\documentclass[12pt,a4paper]{amsart}
\usepackage[utf8]{inputenc}
\usepackage[T1]{fontenc}
\usepackage{times,amsmath,amssymb,color,graphicx,graphics,bm}
\usepackage{url,cite,marginnote}

\newcommand{\bal}{\boldsymbol{\alpha}}
\newcommand{\bbe}{\boldsymbol{\beta}}
\newcommand{\bde}{\boldsymbol{\delta}}

\begin{document}

\title[KdV-type equations for an uneven bottom]{What kinds of KdV-type equations are allowed by an uneven bottom} \thanks{This paper is dedicated to the memory of our friend Professor Eryk Infeld, who recently passed away.
}

\author{Anna Karczewska}
\address{Faculty of Mathematics, Computer Science and Econometrics, University of Zielona G\'ora, Szafrana 4a, 65-246 Zielona G\'ora, Poland.}
\email{A.Karczewska@wmie.uz.zgora.pl}

\author{Piotr Rozmej}
\address{Faculty of Physics and Astronomy, University of Zielona G\'ora, Szafrana 4a, 65-246 Zielona G\'ora, Poland}
\email{P.Rozmej@if.uz.zgora.pl}


\date{\today}

\begin{abstract}
In this study, we give a survey of derivations of KdV-type equations with an uneven bottom for several cases when small (perturbation) parameters $\alpha, \beta, \delta$ are of different orders. Six different cases of such ordering are discussed. Surprisingly, for all these cases the Boussinesq equations can be made compatible only for the particular piecewise linear bottom profiles,  and the correction function has a universal form.  
For such bottom relief, several new KdV-type wave equations are derived. These equations generalize the KdV, the extended KdV (KdV2), the fifth-order KdV (KdV5) and the Gardner equations.

\end{abstract}

\maketitle

\section{Introduction} \label{intro}

The Korteveg de Vries equation (KdV in short) \cite{kdv} belongs to a few most famous equations in mathematical physics. It was originally derived for surface water waves in so-called {\em shallow water wave problem}. In the sixties of the last century, rapid development of the theory of nonlinear waves in various physical systems began, which showed that the KdV equation was obtained as the first approximation in the description of many physical phenomena. The range of applications extends, among others, to waves on the surface of liquids, waves in interfaces between various phases of liquids, ion-acoustic waves in plasma, optical impulses in optical fibers and electrical impulses in electrical circuits. There is a vast number of textbooks and monographs referring to studies of these problems, see, e.g.\ \cite{Whit, DrJ, Ablc, Hir, Rem, InR, Osb}, to list a few. Wonderful properties of the KdV equation like integrability, a rich variety of analytic solutions and the existence of the infinite number of invariants attracted the attention of physicists, mathematicians, and engineers. 

KdV and other KdV-type equations are derived under an important assumption, that the bottom of the fluid container is flat. This assumption is not realistic for most of the situations in the real world, in particular, bottoms of rivers, seas, oceans are non-flat.
Despite a big number of efforts in studying nonlinear waves in the case of a non-flat bottom the first kdV-type equations in which terms originating from the bottom profile occur appeared only recently.
Among the first papers treating a slowly varying bottom are papers by Mei and  Le M\'ehaut\'e~\cite{Mei} and Grimshaw \cite{Grim70}. These authors found that for small amplitudes the wave amplitude varies inversely as the depth but they did not obtain any simple KdV-type equation.
Djordjevi\'c and Redekopp \cite{Djord} and later Benilow and Howlin \cite{BH} studied the motion of packets of surface gravity waves over an uneven bottom using variable coefficient nonlinear
Schr\"odin\-ger equation (NLS).  As a result they found fission of an envelope soliton. Some research groups developed approaches combining linear and nonlinear theories \cite{Pel,Peli,Peli1}.
The Gardner equation  (sometimes called the forced KdV equation) was also extensively investigated \cite{Grim,Smy,Kam,PS98}. 
Van Groeasen and Pudjaprasetya \cite{G&P1,G&P2} applied a Hamiltonian approach in which they obtained a forced KdV-type equation. 
Another widely applied method consists in taking an appropriate average of vertical variables which results in the Green-Naghdi equations \cite{GN,Nad,Kim}. An interesting numerical study of the propagation of unsteady surface gravity waves above an irregular bottom is done in \cite{Sel2000}.
Another study of long wave the propagation over a submerged 2-dimensional bump was recently presented in \cite{NiuYu}, although according to linear long-wave theory. Several examples of recent studies on the propagation of solitary waves over a variable topography are given in \cite{YS2018,YGJC18,FY2018}.

\begin{figure}[b]
\begin{center}
 \resizebox{0.7\columnwidth}{!}{\includegraphics{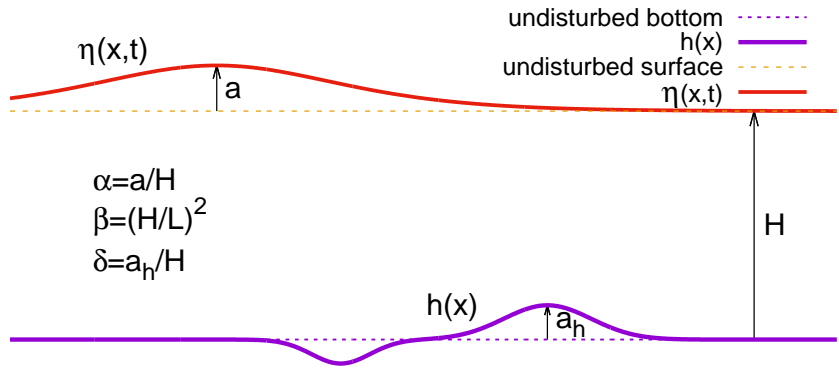}}
 \caption{
Schematic view of the geometry of the shallow water wave problem for an uneven bottom.} \label{geom} 
\end{center} \label{F1}
 \end{figure}

Derivation of wave equations of KdV-type for shallow water problem with an uneven bottom is a difficult task. 
In 2014, with our co-workers, we considered the nonlinear second order wave equation for shallow water problem with uneven bottom \cite{KRR,KRI}. In these papers, besides standard small parameters $\alpha=\frac{a}{H}$ and $\beta=\left(\frac{H}{l}\right)^2$ we introduced the third one defined as $\delta=\frac{a_h}{H}$. In these definitions $a$ denotes the wave amplitude, $H$ the average water depth, $l$ the average wavelength and $a_h$ the amplitude of the variations of the bottom function $h(x)$. The geometry of the considered shallow water problem is presented in Fig.\ \ref{F1}.
Unfortunately, the last step of derivation of the final wave equation for an uneven bottom in \cite{KRR,KRI} was not fully consistent, and therefore the final result is not correct. Below, we present the consistent derivations of such equations, not only for the case when $\alpha,\beta$, and $delta$ are of the same order but also in five different cases of their orderings.

With standard assumptions for incompressible, inviscid fluid and irrotational motion, one obtains the set of Eulerian equations in dimension variables. Next, introduction of the following transformation to dimensionless variables 
\begin{equation}\label{Przeskal}
\tilde{\phi}= \frac{H}{l a \sqrt{gH}}  \phi, \quad  \tilde{x}= x/l, \quad 
\tilde{\eta}= \eta/a, \quad \tilde{z}= z/H, \quad \tilde{t}= t/(l/\sqrt{gH})
\end{equation}
has made it possible to apply perturbation approach, assuming that appropriate parameters are small.

The set of Euler equations, written in nondimensional variables (tildes are now dropped) has the following form (see, e.g., Eqs.\ (2)-(5) in \cite{KRI}) 
\begin{align} \label{2BS}
\beta \phi_{xx}+\phi_{zz} & = 0, \\ \label{4BS}
\eta_t+\alpha\phi_x\eta_x-\frac{1}{\beta}\phi_z & = 0,\quad\mbox{for}\quad  z = 1+\alpha \eta\\  \label{5BS}
\phi_t+\frac{1}{2}\alpha \phi_x^2+\frac{1}{2}\frac{\alpha}{\beta}\phi_z^2 +\eta
-\tau \beta \frac{\eta_{2x}}{(1+\alpha^2\beta\eta_x^2)^{3/2} }
& = 0,\quad\mbox{for}\quad  z = 1+\alpha \eta \\ \label{6BS}
\phi_z-\beta\delta\left( h_x\,\phi_x\right) & =  0,\quad\mbox{for}\quad z=\delta h(x) .
\end{align}
Equation (\ref{2BS}) is the Laplace equation valid for the whole volume of the fluid. Equations (\ref{4BS}) and (\ref{5BS}) are so-called kinematic and dynamic boundary conditions at the surface, respectively. The equation  (\ref{6BS}) represents the boundary condition at the non-flat unpenetrable bottom. 
In (\ref{5BS}), the Bond number $\tau=\frac{T}{\varrho g h^2}$, where $T$ is the surface tension coefficient. For surface gravity waves this term can be safely neglected, since $\tau < 10^{-7}$, but it can be important for waves in thin fluid layers.
For abbreviation all subscripts in (\ref{2BS})-(\ref{6BS}) denote the partial derivatives with respect to particular variables, i.e.\ $\phi_{t}\equiv \frac{\partial \phi}{\partial t}, \eta_{2x}\equiv \frac{\partial^2 \eta}{\partial x^2}$, and so on. 

The velocity potential is seek in the form of power series in the vertical coordinate
\begin{equation} \label{Szer}
\phi(x,z,t)=\sum_{m=0}^\infty z^m\, \phi^{(m)} (x,t),
\end{equation}
where ~$\phi^{(m)} (x,t)$ are yet unknown functions. The Laplace equation (\ref{2BS}) determines $\phi$ in the form which involves only two unknown functions with the lowest $m$-indexes, $f(x,t):=\phi^{(0)} (x,t)$ and $F(x,t):=\phi^{(1)} (x,t)$. Hence,
\begin{equation} \label{Szer1}
\phi(x,z,t)=\sum_{m=0}^\infty \frac{(-1)^m\beta^m}{(2m)!} \frac{\partial^{2m} f(x,t)}{\partial x^{2m}} z^{2m} + \sum_{m=0}^\infty \frac{(-1)^m\beta^{m+1}}{(2m+1)!} \frac{\partial^{2m+1} F(x,t)}{\partial x^{2m+1}} z^{2m+1}.
\end{equation}

The explicit form of this velocity potential reads as
\begin{align} \label{pot8}
\phi & = f-\frac{1}{2}\beta z^2 f_{2x} + \frac{1}{24}\beta^2 z^4 f_{4x} - \frac{1}{720}\beta^3 z^6 f_{6x} + \cdots  
+ \beta z F_x -\frac{1}{6}\beta^2 z^3 F_{3x}+ \frac{1}{120}\beta^3 z^5 F_{5x}
+ \cdots .
\end{align}

Next, one applies the perturbation approach, assuming that parameters $\alpha, \beta, \delta$ are small. As pointed in \cite{BurSerg,Burde}, the proper ordering of small parameters is crucial to obtain appropriate final wave equations. Therefore, for each particular case, the perturbation approach has to be performed separately expressing all parameter by only one (called the \emph{leading parameter}). 
We will be interested in all possible cases of wave equations obtained in the  perturbation approach up to second order. So, we can specify the following cases, see Table \ref{tab1}. 

\begin{table}[hbt] \caption{Different ordering of small parameters considered in the paper.} \label{tab1} \begin{center} 
\begin{tabular}{||c|c|c|c||} \hline \hspace{2ex}Case\hspace{2ex} & \hspace{4ex}$\alpha$\hspace{4ex} & \hspace{4ex}$\beta\hspace{4ex}$ \hspace{4ex}& \hspace{4ex}$\delta$\hspace{4ex}   \\  \hline 
1 & $O(\beta)$ & leading parameter  & $O(\beta)$ \\ \hline 
2 & $O(\beta)$ &  leading parameter & $O(\beta^2)$ \\ \hline
3 & $O(\beta^2)$ &  leading parameter & $O(\beta)$ \\ \hline 
4 & $O(\beta^2)$ & leading parameter  & $O(\beta^2)$ \\ \hline 
5 & leading parameter & $O(\alpha^2)$  & $O(\alpha)$ \\ \hline 
6 & leading parameter & $O(\alpha^2)$  & $O(\alpha^2)$ \\ \hline 
\end{tabular} \end{center}
\end{table}

Our study extends that done thoroughly by Burde and Sergyeyev in \cite{BurSerg}. They considered several cases of a different ordering of two small parameters $\alpha,\beta$, still for the flat bottom case sometimes going up to third or fourth order. The cases studied in \cite{BurSerg} were: $\beta=O(\alpha)$, $\beta=O(\alpha^2)$, $\beta=O(\alpha^3)$, $\alpha=O(\beta^2)$ and  $\alpha=O(\beta^3)$. The authors showed that different ordering of small parameters implied several kinds of wave equations, previously derived in the literature from different physical assumptions. 

The form of velocity potential (\ref{pot8}), determined by (\ref{Szer}) and the Laplace equation (\ref{2BS}), is the same for all considered cases 1-7. The boundary condition at the bottom (\ref{6BS}) implies different forms of the $F$ function, depending on the particular ordering of small parameters $\beta,\delta$.

It is worth noticing the important difference between the cases related to a flat bottom and those when the bottom is not even. In the former ones $F=0$, due to $\delta=0$ in (\ref{6BS}). Therefore, when the Boussinesq equations are used to determine correction terms $Q$ one can always utilize the condition $Q_t=-Q_x$. These facts ensure to derive KdV-type equations up to arbitrary order. 
This is not possible for an uneven bottom. In the latter case the boundary condition (\ref{6BS}) imposes a differential equation on $f$ and $F$ which can be resolved to obtain $F(f(x,t),h(x))$ only in some low orders depending on ordering relations between small parameters. For higher orders that equation cannot be resolved.

In the following sections, we discuss derivations of KdV-type wave equations in cases 1-6. Some examples of numerical simulations illustrating soliton motion over a linearly sloped bottom are presented in Section \ref{num}. The last Section~\ref{concl} contains conclusions.

In this paper, we focused on derivations of wave equations which include terms from the uneven bottom. We leave the broader numerical studies of derived equations to the next article.
 
\section{Case~1: \hspace{1ex}  $\bal=O(\bbe)$, \hspace{1ex} $\bde=O(\bbe)$} \label{cas1}

Due to the velocity potential formula (\ref{pot8}) the natural leading parameter is $\beta$. 
In order to consider perturbation expansion in only one small parameter,  we can set 
\begin{equation} \label{small}
\alpha= A \beta, \quad \delta = D \beta,
\end{equation}
where the constants $A,D$ are of the order of 1. 

Substitution of (\ref{pot8}) into (\ref{6BS}) gives (with $z=D\beta h(x)$) the following nontrivial relation between the functions $F_x$ ~and~$f$ 
\begin{align} \label{fF}
F_{x} & -D\beta (h f_x)_x -\frac{1}{2} D^2\beta^3 (h^2 F_{2x})_x +\frac{1}{6}D^3\beta^4 (h^3 f_{3x})_x 
+\frac{1}{24}D^4 \beta^6 (h^4 F_{4x})_x
+ \cdots=0.
\end{align}
Keeping only terms lower than third order leaves 
\begin{equation} \label{F2}
F_x =D\beta (h f_x)_x,
\end{equation}
which allows us to express the $x$-dependence of the velocity potential through $f,h$ and their $x$-derivatives up to third order. 
With higher order terms in (\ref{fF}) it is impossible. The equation (\ref{F2}) determines $F_x$ up to second order. Since this term enters (\ref{pot8}) with the factor $\beta z$, the velocity potential is determined correctly up to third order in $\beta$.  It is worth to emphasize that due to the presence of the term $-\frac{1}{\beta}\phi_z$ in (\ref{4BS}), the Boussinesq equations resulting from the substitution of (\ref{pot8}) into (\ref{4BS}) and (\ref{5BS}) are correct up to second order. 
Therefore, {\bf the boundary condition at the uneven bottom implies the limit on the order of theory in which the Boussinesq equations can be derived. } For the case $\alpha=O(\beta)$ , $\delta=O(\beta)$ this is second order.


It is easy to see that the form of (\ref{fF}), when considered to higher orders does not allow us for obtaining an explicit expression of $F_x$ through $f,h$ and their $x$-derivatives. 

Substituting (\ref{F2}) into (\ref{pot8}) and retaining terms up to third order in $\beta$ gives the velocity potential as
\begin{align} \label{pot8a}
\phi & = f-\frac{1}{2}\beta z^2 f_{2x} + \frac{1}{24}\beta^2 z^4 f_{4x} 
- \frac{1}{720}\beta^3 z^6 f_{6x} 
+ \beta^2 z D (h f_x)_x -\frac{1}{6}\beta^3 z^3 D (h f_x)_{3x} .
\end{align}
[Due to the term  $\frac{1}{\beta}\phi_z$ in the equation (\ref{4BS}), to obtain equations up to second order one has to keep the velocity potential up to third order.]

Inserting (\ref{pot8a}) into (\ref{4BS}) and (\ref{5BS}), with $z=1+A\beta\eta$,  and retaining terms up to second order yields the set of the Boussinesq equations in the following form (with usual notation $w=f_x$) 
\begin{align} \label{Bus2}
 \eta_t + w_x & + \beta\left(A(\eta w)_x-\frac{1}{6} w_{3x} -D(hw)_x\right) 
\nonumber \\ &  
+\beta^2\left(-A\frac{1}{2} (\eta w_{2x})_ x +\frac{1}{120}w_{5x} + D (hw)_{3x} \right) =0, \\ \label{Bus3} 
w_t + \eta_x & +\beta\left(A w w_x -\frac{1}{2} w_{2xt}- \tau \eta_{3x}\right)
\nonumber \\ &  
+\beta^2\left[A\left(-(\eta w_{xt})_x +\frac{1}{2} w_x w_{2x} - \frac{1}{2} w w_{3x}\right) + \frac{1}{24} w_{4xt}+D(hw_t)_{2x}\right] =0 .
\end{align} 

In the lowest (zero) order the Boussinesq set reduces to 
\begin{align} \label{0ord}
\eta_t + w_x &=0, \quad   w_t + \eta_x =0, \quad \mbox{implying} \quad
 w =\eta, \quad \eta_t =- \eta_x, \quad w_t =-w_x . 
\end{align}
In the first order the Boussinesq set reduces to 
\begin{align} \label{4hx1}
  \eta_t + w_x & + \beta\left( A (\eta w)_x-\frac{1}{6} w_{3x}  -
 D (hw)_x \right) =0,  \\ \label{5hx1}
w_t + \eta_x & + \beta \left(A w w_x -\frac{1}{2} w_{2xt}- \tau \eta_{3x}\right)  =0 . 
\end{align} 
Note that terms originating from an uneven bottom appear in (\ref{4hx1}) but not in (\ref{5hx1}). This is the reason why in first order the Boussinesq equations (\ref{4hx1})-(\ref{5hx1}) can be made compatible only for the particular case of the bottom function $h(x)$.

Assume that in the first order the function $w$ has the form
\begin{equation} \label{w1o}
w= \eta +  \beta\left(-\frac{1}{4}A\, \eta^2+\frac{1}{6}(2-3\tau)\eta_{2x}+D  Q\right).
\end{equation} 
It is easy to see that the first two terms in the correction function assure the KdV equation in the case of the flat bottom.
Then inserting (\ref{w1o}) into (\ref{4hx1})-(\ref{5hx1}) gives in first order
\begin{align} \label{bbus1}
\eta_t + \eta_x & +\beta\left(\frac{3}{2} A \eta\eta_x+\frac{1}{6}(1-3\tau) \eta_{3x} -D (h\eta)_x +D Q_{x}\right) =0 \quad \mbox{and} \\ \label{bbus2}
\eta_t + \eta_x & +\beta\left( \frac{3}{2} A \eta\eta_x+\frac{1}{6}(1-3\tau) \eta_{3x}  +D Q_{t} \right) =0,
\end{align}
where in (\ref{bbus2}) we already replaced $\eta_t$ by $-\eta_x$ 
(from zeroth order). 
The equations (\ref{bbus1}) and (\ref{bbus2}) become compatible when 
\begin{equation} \label{warunek}
 Q_x-Q_t =  (h\eta)_x.
\end{equation}

\subsection{Consequences of the compatibility condition (\ref{warunek}) } \label{comcon}

The same compatibility condition (\ref{warunek}) appears in all cases of the  ordering of small parameters considered in this paper. Now, we discuss the consequences implied by this universal condition.

All known KdV-type equations, for instance the KdV, extended KdV, fifh-order KdV, mKdV, and Gardner equations, have the general form
$ \eta_t = F(\eta,\eta_x,\eta_{2x},\ldots,\eta_{nx}) $
which contains both linear and nonlinear terms. The rhs of the condition (\ref{warunek}) implies that the appropriate correction term $Q$ has to contain, besides $\eta$ and some of its $x$ derivatives, terms with $h$ and its $x$ derivative(s). A term $\int \eta\,dx$ is admissible, as well, since its $x$- and $t$-derivatives are still expressed by local function $\eta$.
Let us seek for $Q$ in the following general form ($a,b$ to be determined)
\begin{equation} \label{genQ}
Q= a\, h\eta +b\,h_x \!\int\! \eta\, dx.
\end{equation}
Then 
\begin{align} \label{qx1r}
Q_x & =a(h_x\eta+ h\eta_x) + b(h_{2x} \!\int\! \eta\, dx+h_x \eta) \quad \mbox{and} \\ \label{qt1r}
Q_t &=ah\eta_t+bh_x \!\int\! \eta_t\, dx= -ah\eta_x+bh_x \!\int\!(- \eta_x)\, dx = -ah\eta_x-bh_x\eta.
\end{align}
In (\ref{qt1r}) we replaced $\eta_t$ by $-\eta_x$ from zeroth order relation.
So, the condition (\ref{warunek}) for $Q$ in the form (\ref{genQ}) is expressed by the formula
\begin{equation} \label{abQ}
(a+2b-1)h_x\eta + (2a-1)\, h\eta_x +bh_{2x} \!\int\! \eta\, dx= 0.
\end{equation}
The equation (\ref{abQ}) is valid only when simultaneously 
$$a+2b-1=0, \quad (2a-1) \quad \mbox{and} \quad bh_{2x} \!\int\! \eta\, dx= 0,
$$
This permits fulfiling the condition (\ref{warunek}), but only for  $h_{2x}=0$, with
$$ a=\frac{1}{2} \quad \mbox{and~ then} \quad  b=\frac{1}{4}.
$$
This important result was first indicated in \cite{Burde}, for cases 1 and 2 of ordering of small parameters. The conclusion in \cite{Burde} stated that the compatibily condition (\ref{warunek}) can be satisfied only for linear bottom function {\bf $\mathbf{h(x)=k\,x}$}, where $k$ is a constant. For this linear bottom function the correction term has the form $Q=\frac{1}{4} (2kx\eta+k\!\int\! \eta dx)$.

The linear function $h=kx$ suffers, however, a significant drawback. For sufficiently large $|x|$ it violates the assumption that $\delta$ is small.
This disadvantage is removed by allowing that 
 $h(x)$ is an arbitrary {\bf piecewise linear} function. Then
\begin{equation} \label{warunQ}
Q= \frac{1}{2} \left(2 h\eta +h_x\!\int\! \eta\, dx\right), \quad \mbox{with} \quad
h_{2x}=0.
\end{equation}

In next sections we will show that this is a universal feature,
necessary for compatibility of the Boussinesq equations for any ordering of small parameters. One has to remember that all newly derived wave equations are valid only for the bottom given by a piecewise linear function.

\subsection{Generalization of KdV for the linear bottom function}

The correction function (\ref{warunQ})
makes the equations (\ref{bbus1}) and (\ref{bbus2}) compatible 
and the resulting first order KdV-type equation has the following form
\begin{equation} \label{kdvD}
\eta_t + \eta_x +\beta\left( \frac{3}{2} A \eta\eta_x+\frac{1}{6}(1-3\tau) \eta_{3x} - \frac{1}{4} D (2 h \eta_x+h_x \eta) \right) =0.
\end{equation}
The equation (\ref{kdvD}) is the new result achieved in \cite{Burde}. 

In original notations for small parameters, this equation reads as
\begin{equation} \label{kdvDo}
\eta_t + \eta_x +\frac{3}{2} \alpha \eta\eta_x+\frac{1}{6}(1-3\tau)\beta \eta_{3x} 
- \frac{1}{4} \delta (2 h \eta_x+h_x \eta) =0.
\end{equation}
For the case of the flat bottom ($D=\delta=0$), (\ref{kdvDo}) reduces to the usual KdV equation
$$\eta_t + \eta_x +\frac{3}{2} \alpha \eta\eta_x+\frac{1}{6}(1-3\tau)\beta \eta_{3x} 
=0.$$
Therefore, the equation (\ref{kdvDo}) generalizes KdV to a case of a piecewise linear bottom profile. 

When $h(x)$ is a bounded arbitrary function the equations (\ref{bbus1})-(\ref{bbus2}) cannot be made compatible. 

Attempts to continue derivation of a wave equation to second order in small parameters show that equations (\ref{Bus2})-(\ref{Bus3}) cannot be made compatible \cite{Burde}. Let us confirm this fact. Assume 
\begin{align} \label{w2o}
w= \eta & +  \beta\left(-\frac{1}{4}A\, \eta^2+\frac{1}{6}(2-3\tau)\eta_{2x}+  
\frac{1}{4} D (2h\eta+h_x\!\int\! \eta dx ) \right) \\ & +\beta^2\left(A^2\frac{1}{8}\eta^3 +A\frac{3+7\tau}{16}\eta_x^2 +A\frac{2+\tau}{4}\eta \eta_{2x}+\frac{12-20\tau-15\tau^2}{120}\eta_{4x} \nonumber
\right)+ \beta^2 D Q
\end{align}
In (\ref{w2o}) we use the form of first order correction (\ref{kdvD}) and the part of second order correction which is appropriate when $D=0$. Therefore $Q$ in (\ref{w2o}) is responsible only for this part which depends on the bottom relief. An attempt to make equations (\ref{Bus2})-(\ref{Bus3}) compatible leads to the condition on $Q_x-Q_t$ which gives no hope for obtaining second order wave equation expressed by local variables $\eta,h$ and their derivatives.

Let us remind that in the case of the flat bottom (with surface tension neglected) the second order wave equation reduces to well known the so-called \emph{extended KdV} equation 
\begin{equation} \label{2kdv} \eta_t + \eta_x +\frac{3}{2} \alpha \eta\eta_x+\frac{1}{6}\beta \eta_{3x} -\frac{3}{8}\alpha^2\eta^2\eta_x +\alpha\beta\left(\frac{23}{24}\eta_x\eta_{2x} +\frac{5}{12}\eta\eta_{3x} \right)+\frac{19}{360}\beta^2\eta_{5x}  =0\end{equation}derived for the first time by Marchant and Smyth in \cite{MS90} and sometimes called \emph{KdV2}.
This equation is nonintegrable. Despite this fact, we with our co-workers found several forms of analytic solutions to KdV2: soliton solutions ($\sim \text{sech}^2[B(x-vt)]$) in \cite{KRI}, cnoidal solutions ($\sim \text{cn}^2[B(x-vt)]$) in \cite{IKRR} and superposition cnoidal solutions ($\sim \text{dn}^2[B(x-vt)]\pm \sqrt{m}\,\text{cn}[B(x-vt)]\,\text{dn}[B(x-vt)]$) in \cite{RKI,RK}.

\section{Case~2: \hspace{1ex} $\bal=O(\bbe)$, \hspace{1ex} $\bde=O(\bbe^2)$}
\label{cas2}

In this case we set
\begin{equation} \label{small2}
\alpha= A\beta, \quad \delta = D \beta^2.
\end{equation}
Now, we insert the general form of velocity potential (\ref{pot8}) into the bottom boundary condition  (\ref{6BS}) which in this case is
\begin{equation} \label{bbc}
\phi_z-D\beta^3\left( h_x\,\phi_x\right) =  0,\quad\mbox{for}\quad z=D\beta^2 h(x) 
\end{equation}
obtaining relation similar to (\ref{fF})
\begin{align} \label{GG2}
F_x & -D\beta^2 (h f_x)_x -\frac{1}{2} D^2\beta^5 (h^2 F_{2x})_x 
 +O(\beta^{7}) =0. 
\end{align} 
From (\ref{GG2}) we have 
\begin{equation} \label{Gd2}
F_x = D \beta^2 (h f_x)_x,
\end{equation}
valid up to fourth order in $\beta$
which inserted into (\ref{pot8}) gives the velocity potential valid up to fourth order
\begin{align} \label{pot16a}
 \phi &= f-\frac{1}{2}\beta z^2 f_{2x} + \frac{1}{24}\beta^2 z^4 f_{4x} - \frac{1}{720}\beta^3 z^6 f_{6x} + D \beta^3 z (h f_x)_x +\frac{1}{40320}\beta^4 z^8 f_{8x} + O(\beta^5)
\end{align}
Therefore, the Boussinesq equations can be consistently derived up to third order (remember term $\frac{1}{\beta}\phi_x$ in (\ref{6BS})). However, we will proceed to second order, only.

Substituting the velocity potential (\ref{pot16a}) into (\ref{4BS})-(\ref{5BS}) and retaining terms up to second order supplies the Boussinesq equations in the following form
\begin{align} \label{4HX}
  \eta_t + w_x &+   \beta\!\left(\!A(\eta w)_x-\frac{1}{6} w_{3x}\!\right)\! 
+\beta^2\!\left(\!-A\frac{1}{2} (\eta w_{2x})_ x +\frac{1}{120}w_{5x} - D\,(hw)_x\!\right)\!  =0, 
\\ \label{5HX}
 w_t + \eta_x & +\beta\!\left(\!A w w_x -\frac{1}{2} w_{2xt}-\tau\eta_{3x}\!\right)\! \\ & 
+  \beta^2\!\left(\!-A(\eta w_{xt})_x +A\frac{1}{2} w_x w_{2x} - A\frac{1}{2} w w_{3x}
+ \frac{1}{24} w_{4xt} \!\right)\! =0.\nonumber
\end{align} 
The Boussinesq equations (\ref{4HX})-(\ref{5HX}) for the the Case 2 and Case 1
are identical when $\delta=D=0$. Since, in (\ref{4HX}) the term $- D\,(hw)_x$ appears only in second order, the first order solutions are those of the KdV,  with 
\begin{equation} \label{h0}
w =  \eta +   \beta\left(-A\frac{1}{4}\eta^2 +\frac{1}{6}(2-3\tau)\eta_{2x} \right), 
\end{equation}
and
\begin{align} 
\eta_t + \eta_x +\beta\left(A\frac{3}{2}\eta\eta_x+ \frac{1}{6}(1-3\tau)\eta_{3x} \right) & =0 \qquad \mbox{or} \nonumber  \\ \label{kdv1}
\eta_t + \eta_x +\alpha \frac{3}{2}\eta\eta_x+ \beta \frac{1}{6}(1-3\tau)\eta_{3x}  & =0 
\end{align}
in original variables.

Now, we aim to satisfy the Boussinesq system (\ref{4HX})-(\ref{5HX}) with the terms of the second order included. Then, we set (the first term with $\beta^2$ is known from the flat bottom case)
\begin{align} \label{wab2}
w = \eta & +\beta\left(-A\frac{1}{4}\eta^2 +\frac{1}{6}(2-3\tau)\eta_{2x} \right)  \nonumber\\ &
 + \beta^2 \left(A^2\frac{1}{8}\eta^3 +A\frac{3+7\tau}{16}\eta_x^2 +A\frac{2+\tau}{4}\eta \eta_{2x}+\frac{12-20\tau-15\tau^2}{120}\eta_{4x} \right)
+ \beta^2 D Q.
\end{align}
Next, we insert the trial function (\ref{wab2}) into (\ref{4HX}) and (\ref{5HX}) and retain terms up to second order in $\beta$. Proceeding analogously as in the case of first order we find that compatibility of the Boussinesq equations (\ref{4HX})-(\ref{5HX})  requires the following condition for the correction function $Q$
$$ Q_x- Q_t =  (h \eta)_x,$$
the same as the condition (\ref{warunek}).

Note, that in order to replace $t$-derivatives by $x$-derivatives one has to use  the properties of the first order equation (\ref{kdv1}), that is,  $ \eta_t= -\eta_x-\beta\left( A\frac{3}{2}\eta\eta_x +\frac{1-3\tau}{6}\eta_{3x}\right)$ and its derivatives.

Using universal formulas for the correction functions, obtained for a piecewise linear bottom  
we obtained in this case, $\alpha=O(\beta), \delta=O(\beta^2)$, the equation 
\begin{align}\label{2abQ}
\eta_t+\eta_x +\frac{3}{2}\alpha\eta\eta_x  +\frac{1-3\tau}{6}\beta\eta_{3x} &
-\frac{3}{8}\alpha^2\eta^2\eta_x +\alpha\beta\left(\frac{23+15\tau}{24}\eta_x^2  + \frac{5-3\tau}{12}\eta\eta_{2x} \right)\nonumber \\ & + \beta^2\left(\frac{19-30\tau-45\tau^2}{360}\eta_{5x} \right)  -\frac{1}{4}\delta (2 h \eta_x+h_x \eta) =0 \end{align}
which generalizes the extended KdV (KdV2) equation (\ref{2kdv}).

\section{Case 3: \hspace{1ex} $\bal=O(\bbe^2)$, \hspace{1ex} $\bde=O(\bbe)$}
\label{cas3}

In this case we set
\begin{equation} \label{small3}
\alpha= A\beta^2, \quad \delta = D \beta.
\end{equation}

Since $\delta$ is of the same order as $\beta$ the formulas (\ref{fF})-(\ref{pot8a}) expressing the velocity potential hold.
Now we substitute the velocity potential (\ref{pot8a}) into the kinematic and dynamic boundary conditions at the unknown surface which in this case are
\begin{eqnarray}  \label{4bs}
\eta_t+A\beta^2\phi_x\eta_x-\frac{1}{\beta}\phi_z & = &0, \quad \mbox{for} \quad   z = 1+A\beta^2 \eta ,\\  \label{5bs}
\phi_t+\frac{1}{2}A\beta^2 \phi_x^2+\frac{1}{2}A\beta \phi_z^2 +\eta 
-\tau \beta \frac{\eta_{2x}}{(1+A^2\beta^5\eta_x^2)^{3/2} }
& = &0,\quad \mbox{for} \quad  z = 1+A\beta^2 \eta  .
\end{eqnarray}
Next, we neglect all terms of orders higher than $\beta^2$. The result consists in the following Boussinesq equations (in the meantime the second equation was differentiated by $x$)
\begin{align} \label{r4BS} 
\eta_t+w_x & -\beta\left(D\, (h w)_x +\frac{1}{6}w_{3x} \right)+\beta^2 \left( A(\eta w)_x + \frac{1}{2}D (h w)_{3x}+ \frac{1}{120}w_{5x} \right)  = 0 ,\\ 
\label{r5BS}
w_t+\eta_x & - \beta \left( \tau\eta_{3x} + \frac{1}{2} w_{2xt} \right) +\beta^2 \left( A w w_x +D (h w_{t})_{2x}+  \frac{1}{24}w_{4xt} \right) = 0 ,
\end{align}
where the usual notation ~$w=f_x$~ is used. 

For the flat bottom case ($D=0$) the equations (\ref{r4BS})-(\ref{r5BS}) can be made compatible up to any order. Some of them are derived from different physical models.
Below we cite this equations keeping terms up to second order in $\beta$ (see, \cite[Eqs.~(A.1)-(A.2)]{BurSerg})
\begin{align} \label{wBS2}
& w = \eta +\beta \frac{2-3\tau}{6}\eta_{2x} +\beta^2\left(-A\frac{1}{4}\eta^2+\frac{12-20\tau-15\tau^2}{120}\eta_{4x} \right), \\  \label{eWS2}
& \eta_t +\eta_x + \beta \frac{1-3\tau}{6}\eta_{3x}+\beta^2\left(A\frac{3}{2}\eta\eta_x+\frac{19-30\tau-45\tau^2}{360}\eta_{5x} \right) =0.
\end{align}
This result is equivalent to well known \emph{the fifth-order KdV equation} derived
by Hunter and Sheurle in \cite{HS88} as a model equation for gravity-capillary shallow water waves of small amplitude. 
Neglecting surface tension (reasonable for shallow water problem) and changing variables by 
$$ \tilde{x}=\sqrt{\frac{3\alpha}{2\beta}}\,(x-t),\qquad \tilde{t} =\frac{1}{4}\sqrt{\frac{3\alpha^3}{2\beta}}\,t ,$$
 one reduces the equation (\ref{eWS2}) to 
$$\eta_{\tilde{t}}+6\eta\eta_{\tilde{x}}+\eta_{3\tilde{x}} + P\, \eta_{5\tilde{x}}=0, \qquad P = \frac{19}{40},
$$
which is the fifth-order KdV equation obtained in \cite{HS88} with $P$ defined in a different way. This equation is known to have a rich structure of solitary wave solutions, see, e.g., \cite{GMB94}. 
As pointed out in \cite{BurSerg} the wave equation obtained 
in third order belongs to the type $K(m,n)$ introduced by Rosenau and Hyman \cite{RH93} with $m=4$ and $n=1$ which in some range of wave velocities admits soliton-like traveling wave solutions.

For the case $D\ne0$ limitation of equations (\ref{r4BS})-(\ref{r5BS}) to first order yields
\begin{equation} \label{r45BS}
\eta_t+w_x-\beta\left(D\, (h w)_x-\frac{1}{6}w_{3x} \right)  = 0 \qquad \mbox{and} \qquad
\eta_x+w_t -\beta\left(\tau\eta_{3x}+ \frac{1}{2}  w_{2xt}\right)  = 0 .
\end{equation}
Since in zeroth order $\eta=w$, $\eta_t=-\eta_x$, $w_t=-w_x$, one assumes that in the first order
\begin{equation} \label{r45w}
w=\eta+\beta \left(\frac{2-3\tau}{6}\eta_{2x} + D Q\right) ,
\end{equation}
where the first part of the correction term is alraedy known from (\ref{wBS2}) and $Q$ is responsibe for first order correction related to the bottom term in (\ref{r45BS}). Then, substitute (\ref{r45w}) into equations (\ref{r45BS}) and retain terms only to the first order. This yields
\begin{equation} \label{r45qxt}
\eta_t+\eta_x+\beta\left(D Q_x- D\, (h \eta)_x +\frac{1-3\tau}{6}\eta_{3x} \right)=0
\qquad \mbox{and} \qquad \eta_t+\eta_x+\beta\left(D Q_t+\frac{1-3\tau}{6}\eta_{3x}\right)=0.
\end{equation}
Compatibility of these equations requires
\begin{equation} \label{rQxt}
Q_x-Q_t =  (h\eta).
\end{equation}
This is the same condition as (\ref{warunek}) which cannot be satisfied for general 
form of $h(x)$ but can be satisfied for the {\bf particular case $\mathbf{h(x)=k\,x}$.}
In this particular case $Q$ given by (\ref{warunQ}) makes the equations (\ref{r45qxt}) compatible. So, with 
\begin{equation} \label{wD0}
w= \eta+\beta \left(\frac{2-3\tau}{6}\eta_{2x} +\frac{1}{4} D 
\left(2h\eta +h_x\!\int\! \eta\,dx\right) \right)
\end{equation}
 we obtain the resulting first order KdV-type equation in the following form
\begin{equation} \label{kdvDa}
\eta_t + \eta_x +\beta\left(\frac{1-3\tau}{6}\, \eta_{3x} - \frac{1}{4} D 
(2 h \eta_x+h_x \eta) \right) =0.
\end{equation}

Note, that the equation (\ref{kdvDa}), in the case of the flat bottom ($D=0$), is reduced to the linear dispersive one. Therefore the equation (\ref{kdvDa}) has no soliton solutions.

Can we derive a reasonable second order equation? 
Let us assume 
\begin{align} \label{w5rz2}
w= \eta &+\beta \left(\frac{2-3\tau}{6}\eta_{2x} +\frac{1}{4} D \left(2h\eta +h_x\!\int\! \eta\,dx\right) \right)\\ &+\beta^2\left(-A\frac{1}{4}\eta^2+\frac{12-20\tau-15\tau^2}{120}\eta_{4x} + D Q\right). \nonumber
\end{align}
In (\ref{w5rz2}) we already used the form of the first order correction (\ref{wD0}) and second order correction for flat bottom case (\ref{wBS2}). Now, we seek for $Q$ which is the second order correction originating only from the bottom terms in the Boussinesq equations (\ref{r4BS})-(\ref{r5BS}).
Inserting (\ref{w5rz2}) into (\ref{r4BS})-(\ref{r5BS}) and replacing $t$-derivatives by $x$ derivatives according to (\ref{kdvDa}) we obtain the following condition for the second order correction $Q$
\begin{align} \label{Q2rz}
Q_x-Q_t = &~ -k\left(\frac{1}{4}(\eta+x\eta_x) +\frac{1+\tau}{2}(\eta_{2x}+ x\eta_{3x})\right)+k^2 D\left(\frac{11}{16}x\eta+\frac{3}{8}x^2\eta_x  \right) \\  &~ +k^2 D \left(\frac{5}{32}\!\int\!\eta dx + \frac{1}{16}\!\int\!x\eta_x dx\right) \nonumber
\end{align}
This condition seems to have no solutions which could supply the second order wave equation of similar form  as the first order equation (\ref{kdvDa}), that is, $\eta_t=-[\eta_x +F(\eta,\eta_x,...,x)]$.

\section{Case 4: \hspace{1ex} $\bal=O(\bbe^2)$,\hspace{1ex} $\bde=O(\bbe^2)$}
\label{cas4}

In this case we set
\begin{equation} \label{small4}
\alpha= A \beta^2, \quad \delta = D \beta^2.
\end{equation}
Since $\delta =O(\beta^2)$, the forms of the function $F$ and the velocity potential are given by (\ref{Gd2})-(\ref{pot16a}). The Boussinesq set receives in this case the following form
\begin{align} \label{3Ba2d2}
\eta_t+w_x &-\beta\frac{1}{6} w_{3x}+\beta^2 \left(A(w\eta)_x +\frac{1}{120} w_{5x}-D(h w)_x\right)=0,\\ \label{4Ba2d2}
w_t+\eta_x & -\beta\left(\frac{1}{2} w_{2xt}+\tau \eta_{3x}\right) +\beta^2\left(A w w_x+ \frac{1}{24}w_{4xt}\right)= 0.
\end{align}
In first order $w$ in the form 
\begin{equation} \label{wrz0}
w = \eta +\beta \frac{2-3\tau}{6} \eta_{2x}
\end{equation}
makes the equations (\ref{3Ba2d2})-(\ref{4Ba2d2}) compatible, with the result which is the linear equation
\begin{equation} \label{rrz1}
\eta_t+\eta_x +\beta \frac{1-3\tau}{6} \eta_{3x} =0.
\end{equation}
In second order we look for $w$ in the form
\begin{equation} \label{wrz1}
w = \eta +\beta \frac{2-3\tau}{6} \eta_{2x} +\beta^2 \left(-\frac{1}{4} A\,\eta^2
+\frac{12-20\tau-15\tau^2}{120}\eta_{4x}+D Q\right).
\end{equation}
In (\ref{wrz1}) we already used the part of second order correction term known to make compatible Boussinesq's set for the flat bottom (see, e.g., \cite[Eqs.~(A.8)-(A.9)]{BurSerg}. 
Substitution of (\ref{wrz1}) into (\ref{3Ba2d2})-(\ref{4Ba2d2}) gives  
\begin{align} \label{r55}
\eta_t+\eta_x & +\beta \frac{1-3\tau}{6} \eta_{3x} +\beta^2 \left(D Q_x-D(h\eta)_x+\frac{3}{2} A\eta\eta_x+\frac{19-30\tau-45\tau^2}{360}\eta_{5x}  \right) =0 \quad \mbox{and}\\ \label{r56}
\eta_t+\eta_x & +\beta \frac{1-3\tau}{6} \eta_{3x} +\beta^2 \left(D Q_t + \frac{3}{2} A\eta\eta_x+\frac{19-30\tau-45\tau^2}{360}\eta_{5x} \right)=0, 
\end{align}
respectively. (In (\ref{r56}) $t$-derivatives are already properly replaced by $x$-derivatives from first order equation (\ref{rrz1}.)
Compatibility of equations (\ref{r55})-(\ref{r56}) requires the same condition as (\ref{warunek})
$$Q_x- Q_t=  (h\eta)_x.$$

Using the correction $Q$ given by (\ref{warunQ}) we obtain the wave equation  (in original parameters)
\begin{equation}\label{5kdvQ}
\eta_t+\eta_x + \frac{3}{2}\alpha\eta\eta_x +\beta \frac{1-3\tau}{6} \eta_{3x} +\beta^2 \frac{19-30\tau-45\tau^2}{360}\eta_{5x} -\frac{1}{4} \delta 
(2 h \eta_x+h_x \eta) =0.
\end{equation}
This equation is the generalization of the fifth-order KdV equation (\ref{eWS2}) for the case of uneven bottom. However, the equation (\ref{5kdvQ}) is correct only for a piecewise linear bottom profile.

\section{Case 5: \hspace{1ex}  $\bbe=O(\bal^2)$, \hspace{1ex} $\bde=O(\bal)$}
\label{cas5}

In this case the leading parameter is $\alpha$.  We set
\begin{equation} \label{small5}
\beta= B\alpha^2, \quad \delta = D \alpha.
\end{equation}
Now, we have to express all perturbation equations with respect to parameter $\alpha$. Then the velocity potential (\ref{pot8}) can be rewritten as
\begin{align} \label{po8}
\phi  = f & -\frac{1}{2}B \alpha^2 z^2 f_{2x} + \frac{1}{24}B^2 \alpha^4 z^4 f_{4x} - \frac{1}{720}B^3 \alpha^6 z^6 f_{6x} + \cdots \\ & 
+ B \alpha^2 z F_x -\frac{1}{6}B^2 \alpha^4 z^3 F_{3x}+ \frac{1}{120}B^3 \alpha^6 z^5 F_{5x} + \cdots . \nonumber 
\end{align}
The boundary condition at the bottom (\ref{6BS}) takes now the following form
\begin{equation} \label{bbc1}
\phi_z-BD\alpha^3\left( h_x\,\phi_x\right)  =  0 \quad\mbox{for}\quad z=D\alpha\, h(x).
\end{equation}
Applying this equation to $\phi$ given by (\ref{po8}) implies
\begin{equation} \label{FxAll}
F_x = \alpha D (h f_x)_x + \alpha^4 B D^2(h^2F_{2x})_x - \alpha^5 \frac{1}{6} B D^3(h^3 f_{3x})_x + O(\alpha^8) .
\end{equation}
This equation allows us to express $F_x$ through $h,f$ and their derivatives only when terms of the fourth and higher orders are neglected
\begin{equation} \label{Fx}
F_x = \alpha D (h f_x)_x + O(\alpha^4). 
\end{equation}
This formula allows us to express $\phi$ through only one unknown function $f$ and its derivatives. Note that next terms in $F_x$ would enter in $\phi$ in 
sixth order.
Therefore we can express the velocity potential containing terms from the bottom function only up to fifth order in $\alpha$
\begin{align} \label{po8a}
\phi  = f & -\frac{1}{2}B \alpha^2 z^2 f_{2x} + \frac{1}{24}B^2 \alpha^4 z^4 f_{4x} 
+ B D \alpha^3 z (h f_x)_x  -\frac{1}{6}B^2 D \alpha^5 z^3 (h f_x)_{3x} 
+ O(\alpha^6).
\end{align}
This form of the velocity potential implies the Boussinesq set as
\begin{align} \label{bu3}
\eta_t + w_x  & + \alpha\,[ (\eta w)_x-D (h w)_x]-\frac{1}{6} \alpha^2 B\, w_{3x} +\alpha^3 B\frac{1}{2}\,[D (h w)_{3x}- (\eta w_{2x})_x ]=0, \\ \label{bu4}
w_t + \eta_x & +\alpha\,w w_x -\alpha^2 B\,[ \tau \eta_{3x}+\frac{1}{2} w_{2xt}] \\ &
+\alpha^3 B\,[-(\eta w_{xt})_x + \frac{1}{2}\,w_x w_{2x}- \frac{1}{2}\,w w_{3x}+ D  (h w_t)_{2x}] =0, \nonumber
\end{align}
where terms up to $\alpha^3$ are retained. Note that for the uneven bottom ($\delta,D\ne 0$) consistent  Boussinesq's set cannot be derived in orders higher than $\alpha^3$. Due to the term $\phi_z/\beta=\phi_z/(B\alpha^2)$ in the kinematic boundary condition at the surface (\ref{4BS}) and relations (\ref{FxAll})-(\ref{Fx}) we cannot extend equations (\ref{bu3})-(\ref{bu4}) with the potential (\ref{po8a}) to higher orders. When the bottom is flat, there is no such limitation.  

Begin with first order Boussinesq's set. Assuming $w=\eta+\alpha(-\frac{1}{4} \eta^2)+\alpha D Q$ one obtains from (\ref{bu3})-(\ref{bu4})
\begin{align} \label{b3_1} 
\eta_t + \eta_x  & + \alpha\left[\frac{3}{2}\eta\eta_x + D Q_x  -D (h\eta)_x\right]=0, \\ \label{b3_2}
\eta_t + \eta_x  & + \alpha\left[\frac{3}{2}\eta\eta_x +D Q_t \right] =0 .
\end{align}
Equations (\ref{bu3})-(\ref{bu4}) become compatible in first order when the condition (\ref{warunek}) holds, that is
$$ 
Q_x -Q_t =  (h\eta)_x .
$$ 

Using the correction $Q$ given by (\ref{warunQ}) we obtain first order equation for the piecewise linear bottom relief as
\begin{equation} \label{kdvDb}
\eta_t + \eta_x +\frac{3}{2} \alpha \eta\eta_x - \frac{1}{4} \delta 
 (2 h \eta_x+h_x \eta) =0.
\end{equation}
This equation does not contain the dispersive term $\eta_{3x}$.

An attempt to derive second order wave equation fails. We can assume
\begin{equation} \label{ab2dQ2}
 w=\eta+\alpha(-\frac{1}{4} \eta^2)+\alpha D \frac{1}{4} \left(2h\eta_x+ h_x\!\int\!\eta\, dx\right)+ \alpha^2 \frac{2-3\tau}{6}\eta_{2x} +\alpha^2 D \,Q.
\end{equation}
In $w$ given by (\ref{ab2dQ2}) three first terms ensure compatibility of the Boussinesq equations (\ref{bu3})-(\ref{bu4}) in first order, the fourth one makes terms with $\alpha^2 B=\beta$ compatible. Then the only unknown part of correction 
gives a condition on $Q_x-Q_t$
which does not give any hope for determining $Q$ through $\eta,\eta_x,\ldots,h,h_x$  and then deriving second order wave equation.

\section{Case 6: \hspace{1ex} $\bbe=O(\bal^2)$, \hspace{1ex} $\bde=O(\bal^2)$}
\label{cas6}

The leading parameter is $\alpha$. We set
\begin{equation} \label{small6}
\beta= B\alpha^2, \quad \delta = D \alpha^2.
\end{equation}
The velocity potential is expressed, as in the section \ref{cas5}, by (\ref{po8}), but the boundary condition at the bottom (\ref{6BS}) takes now the form
\begin{equation} \label{bbc2}
\phi_z-BD\alpha^4\left( h_x\,\phi_x\right)  =  0,\quad\mbox{for}\quad z=D\alpha^2 h(x).
\end{equation}
So, from (\ref{po8}) and (\ref{bbc2}) one gets
$$F_x = \alpha^2 D (hf_x)_x + \frac{1}{2} \alpha^6 B D^2 (h^2 F_{2x})_x - \frac{1}{6} \alpha^8 B D^3 (h^3 f_{3x})_x + O( \alpha^{12}) =0.
$$
Neglecting higher order terms we can use 
\begin{equation} \label{Fx2}
F_x = \alpha^2 D (hf_x)_x +O( \alpha^{6}),
\end{equation}
which ensures the expression of $\phi$ through only one unknown function $f$ and its derivatives. Note that next terms in $F_x$ would enter in $\phi$ in $\alpha^8$ order.
Then we can express the velocity potential as
\begin{align} \label{po8b}
\phi  = f & -\frac{1}{2}B \alpha^2 z^2 f_{2x} + \frac{1}{24}B^2 \alpha^4 z^4 f_{4x} - \frac{1}{720}B^3 \alpha^6 z^6 f_{6x} 
\\ & 
+ B D \alpha^4 z (h f_x)_x  -\frac{1}{6}B^2 D \alpha^6 z^3 (h f_x)_{3x} 
+ O(\alpha^8).\nonumber 
\end{align}
With potential given by (\ref{po8b}) we obtain from (\ref{4BS}) and (\ref{5BS}) the Boussinesq set
\begin{align} \label{bbu3}
\eta_t + w_x  & + \alpha (\eta w)_x - \alpha^2\left(\frac{1}{6} B w_{3x}+D (h w)_x \right) -\alpha^3 \frac{1}{2} B (\eta w_{2x})_x \\ & 
+ \alpha^4 \left(\frac{1}{2}B D (h w)_{3x} 
- \frac{1}{2}B (\eta^2 w_{2x})_x + \frac{1}{120} B^2 w_{5x}
\right)=0, \nonumber \\  \label{bbu4}
w_t + \eta_x & + \alpha w w_x -\alpha^2 B \left(\tau \eta_{3x}+  \frac{1}{2} w_{2xt}
\right) + \alpha^3 B \left(-(\eta w_{xt})_x + \frac{1}{2} w_x w_{2x} - \frac{1}{2} w w_{3x}\right) \\ & 
+ \alpha^4 B \left( D (h w_t)_{2x} + w_x(\eta w_x)_x  
-w (\eta w_{2x})_x -\frac{1}{2} (\eta^2 w_{xt})_x +
\frac{1}{24} B w_{4xt} \right) =0, \nonumber 
\end{align}
respectively, where terms up to $\alpha^4$ are retained. 

In first order  
$w=\eta + \alpha (-\frac{1}{4} \eta^2)$  makes (\ref{bbu3})-(\ref{bbu4}) compatible giving
\begin{equation} \label{1rz6}
\eta_t+\eta_x +\frac{3}{2} \alpha \eta\eta_x=0.
\end{equation}
In second order one assumes $w$ in the form 
\begin{equation} \label{wgar}
w = \eta -\frac{1}{4}\alpha \eta^2 + \alpha^2\left(\frac{1}{8}\eta^3 +\frac{2-3\tau}{6}\eta_{2x} \right)+\alpha^2 D Q ,
\end{equation}
which substituted to (\ref{bbu3})-(\ref{bbu4}) gives 
(after neglection of terms of higher orders) 
\begin{equation} \label{2rz3}
\eta_t+\eta_x +\frac{3}{2} \alpha  \eta\eta_x +\alpha^2 \left(- \frac{3}{8} \eta^2\eta_x - \frac{1-3\tau}{6} B \eta_{3x} -D (h\eta)_x +D Q_x\right)=0 
\end{equation}
from (\ref{bbu3}) and
\begin{equation} \label{2rz4}
\eta_t+\eta_x +\frac{3}{2} \alpha  \eta\eta_x +\alpha^2 \left(- \frac{3}{8} \eta^2\eta_x - \frac{1-3\tau}{6} B \eta_{3x} +D Q_t\right)=0 
\end{equation}
from (\ref{bbu4}). Compatibility of equations (\ref{2rz3}) and (\ref{2rz4}) requires the same condition as (\ref{warunek})
$$  Q_x-Q_t= (h\eta)_x.$$

For a piecewise linear bottom profile $h=kx$, due to the universal result (\ref{warunek}) given in subsection \ref{comcon} we obtain 
\begin{equation} \label{GardH}
\eta_t+\eta_x +\frac{3}{2} \alpha  \eta\eta_x +\alpha^2 \left(- \frac{3}{8} \eta^2\eta_x - \frac{1-3\tau}{6} B \eta_{3x} \right)-\frac{1}{4}\delta (2 h \eta_x+h_x \eta) =0 .
\end{equation}

Note that for the flat bottom ($D=0$), $w$ given by (\ref{wgar}) makes the Boussinesq equations (\ref{bbu3})-(\ref{bbu4}) compatible and gives the well known Gardner equation
\begin{equation} \label{Gard}
\eta_t+\eta_x +\frac{3}{2} \alpha  \eta\eta_x +\alpha^2 \left(- \frac{3}{8} \eta^2\eta_x - \frac{1-3\tau}{6} B \eta_{3x} \right)=0 .
\end{equation}

Therefore, the equation (\ref{GardH}) is the generalization of the Gardner equation for the case of the uneven bottom, valid when the bottom is given by the piecewise linear function.

\section{Examples of numerical solutions} \label{num}

In this section, we show some examples of motion of solitons, initialy moving over an even bottom when they enter the region where the bottom is no longer even. These results are obtained by numerical time evolution according to wave equations derived in previous sections. 
In order to be able to compare the influence of the bottom on the soliton movement, in all cases presented in this section, we assumed the same bottom shape in the form of a piecewise linear function. We discuss here only three cases: Case~1, where KdV solitons exist, Case~2, where KdV2 solitons were discovered by us in \cite{KRI} and Case~4, where KdV5 solitons are known \cite{Dey96,Bri02}, as well.    

In the calculations we use our finite difference code described in detail in \cite{KRI}. The cases~1 and~2 in which  $\alpha=O(\beta)$ are appropriate for shallow water waves where surface tension can be safely neglected. This is because when average water depth is of the order of some meters then $\tau< 10^{-7}$. Therefore in the simulations presented in subsections \ref{cas1n} and \ref{cas2n} we set $\tau=0$ in the corresponding wave equations.

\subsection{{\bf Case 1}} \label{cas1n}
In this case $\alpha=O(\beta),~ \delta=O(\beta)$, we performed calculations according to the first order equation (\ref{kdvD}). The piecewise linear bottom function is taken in the form 
\begin{equation} \label{shapeB}
h(x)=\left\{\begin{array}{rcc} 0 &\quad \mbox{for}\quad & x\le 0, \\ \frac{1}{X}\,x &\quad\mbox{for} \quad& 0< x\le X\\ 1 &\quad\mbox{for}\quad & x>X  ~,\end{array}\right. 
\end{equation}
with $X=15$ for all cases considered. 
The shape of the bottom is displayed in presented figures below the snapshots of soliton's motion, not in the same scale.

\begin{figure}[hbt]
\begin{center}
 \resizebox{0.8\columnwidth}{!}{\includegraphics[angle=270]{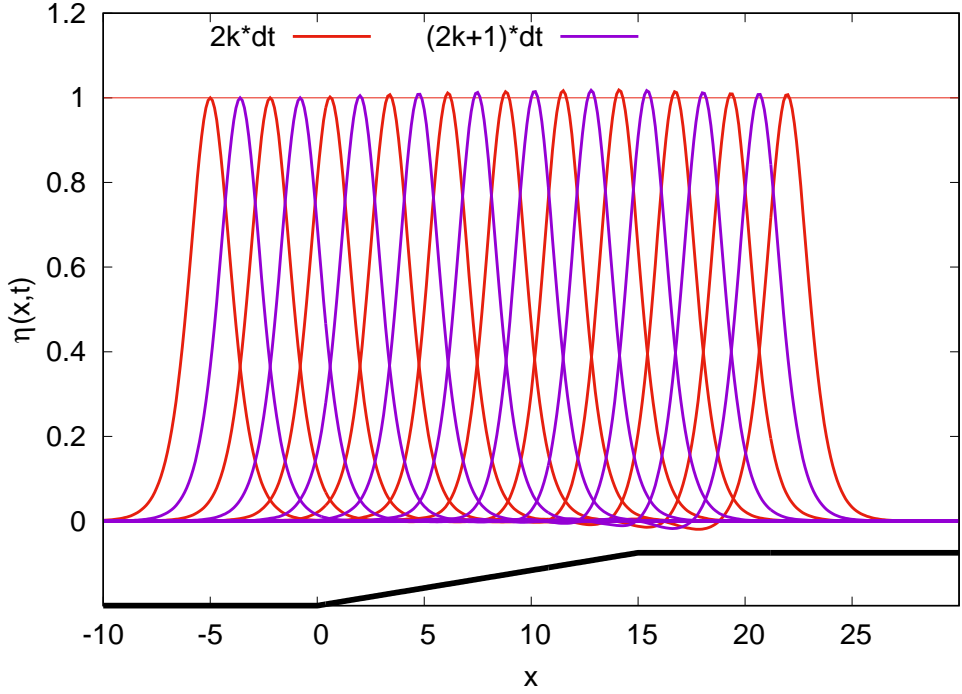}}
 \caption{
Time evolution of the KdV soliton entering the region with the piecewise linear bottom function (\ref{shapeB}) obtained in numerical inegration of the equation (\ref{kdvD}) with parameters $\alpha=0.2424, ~\beta=0.3, ~\delta=0.15$. } \label{Fi2} 
\end{center} 
 \end{figure}

In Fig.~\ref{Fi2}, 
we dispalyed the results of numerical simulation obtained for the following parameters: $\alpha=0.2424$, $\beta=0.3$, $\delta=0.15$.
As the initial condition we took the KdV soliton
\begin{align} \label{1skdv}
\eta(x,t)&=A\,\text{sech}^2\left[B(x-x_0-vt)\right], 
\quad \mbox{where} \quad B= \sqrt{\frac{3\alpha}{4\beta}A\;}\quad \mbox{and} \quad v=1+\frac{\alpha}{2}A, 
\end{align}
with $x_0=-5$, $t=0$ and the amplitude  $A=1$ (to compare with the Case~2). 

If we come back to dimension variables then the soliton's amplitude in meters is obtained by multiplying its nondimenional value by $H$.
For instance, if $H=2\,m$ then the initial soliton's height (above the undisturbed wate level) is $H \alpha \approx 0.485\, m$.
 The horizontal coordinates have to be multiplied by $l=\frac{H}{\sqrt{\beta}}\approx 3.65\,m$.
So, the range $x\in[0,15]$ in the displayed figures corresponds to the interval (approximately) $[0,54]$ in meters. The time increment between the consecutive profiles in presented Figs.~\ref{Fi2} and \ref{Fi3} is $dt=1.25$ (in dimensionless units) what corresponds to approximately $0.56\,s$. Note that $v_0
\approx 4.96\,m/s$ . 

\subsection{{\bf Case 2}} \label{cas2n}

\begin{figure}[hbt]
\begin{center}
 \resizebox{0.8\columnwidth}{!}{\includegraphics[angle=270]{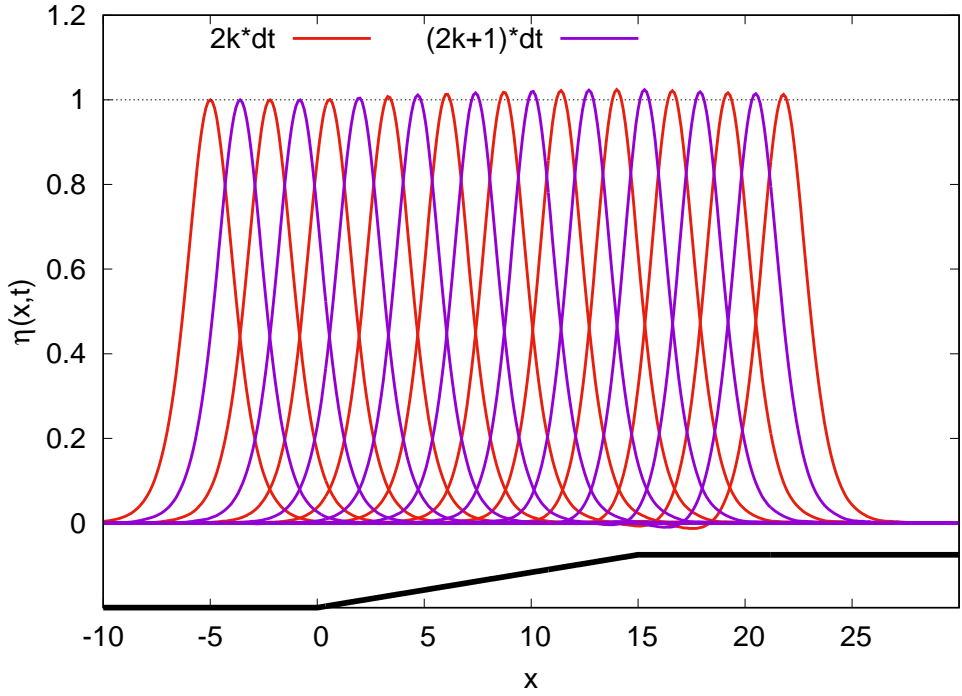}}
 \caption{
Time evolution of the KdV2 soliton entering the region with the piecewise linear bottom function (\ref{shapeB}) obtained in numerical inegration of the equation (\ref{2abQ}) with parameters $\alpha=0.2424, ~\beta=0.3, ~\delta=0.15$. } \label{Fi3}
\end{center} 
 \end{figure}

In this case, $\alpha=O(\beta),~ \delta=O(\beta^2)$, the appropriate wave equation is the equation (\ref{2abQ}). It is worth to remind that when $\delta=0$, the equation (\ref{2abQ}) reduces to the extended KdV (KdV2). 
Since KdV2 possesses exact soliton solution \cite{KRI}, we use this solution as the initial condition in the example presented in Fig.~\ref{Fi3}.

Contrary to the KdV equation which leaves one parameter freedom for the coefficients of the exact solutions (therefore KdV permits for solitons of different amplitudes), the parameters $\alpha,\beta$ of the KdV2 equation
fix the coefficients of the unique soliton solution. So, for the evolution shown in Fig.~\ref{Fi3} the initial condition has the the same form (\ref{1skdv})  but with coefficients: $A\approx \frac{0.2424}{\alpha},~B\approx\sqrt{0.6\,\frac{\alpha}{\beta}A}$ and $v\approx 1.11455$. The parameter $\alpha=0.2424$ assures the amplitude equal one.

For comparison of the KdV2 soliton motion according to the equation (\ref{2abQ}) to the KdV soliton motion according to the equation (\ref{kdvD}) we used the same values of the parameters $\alpha,\beta,\delta$.  Using $\delta=0.15= D\,\beta^2$ with $D\approx 1.67$, not much different from unity, does not contradict to the assumption $\delta=O(\beta^2)$.

\subsection{{\bf Case 4}} \label{cas4n}

\begin{figure}[hbt]
\begin{center}
 \resizebox{0.8\columnwidth}{!}{\includegraphics[angle=270]{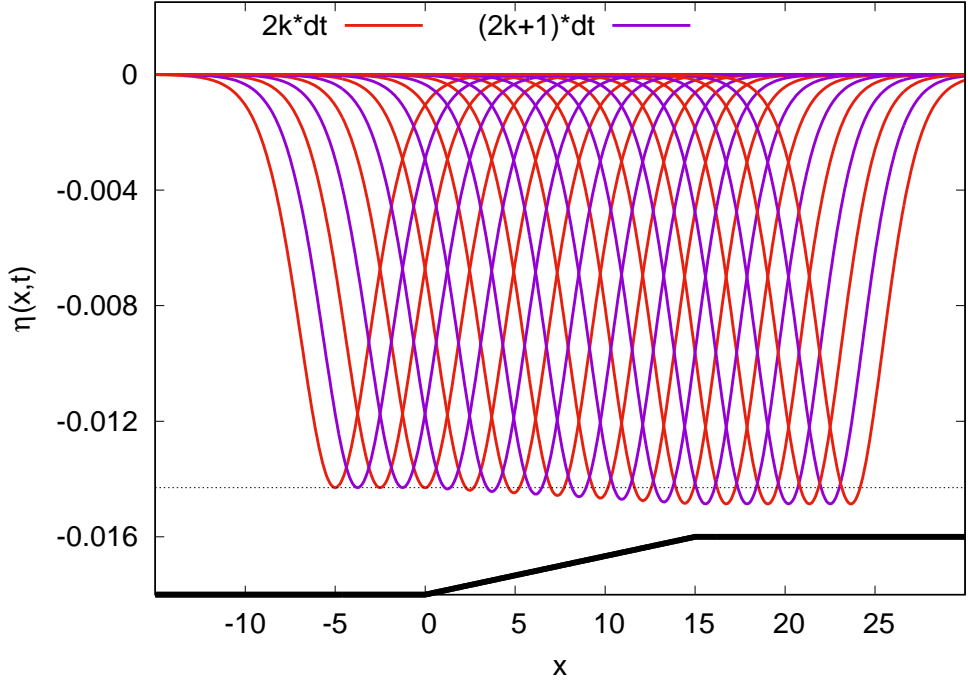}}
 \caption{
Time evolution of the fifth order KdV soliton entering the region with the  piecewise linear bottom function (\ref{shapeB}) obtained in numerical inegration of the equation (\ref{5kdvQ}) with parameters $\alpha=0.2424, ~\beta=0.3, ~\delta=0.15$. } \label{Fi4}
\end{center} 
 \end{figure}

In this case, $\alpha=O(\beta^2),~ \delta=O(\beta^2)$, the appropriate wave equation is the equation~(\ref{5kdvQ}). It is worth to remind that when $\delta=0$, the equation (\ref{5kdvQ}) reduces to (\ref{eWS2}), the \emph{fifth-order KdV} or \emph{KdV5} \cite{HS88,GMB94,Dey96,Bri02}. 
The fifth-order KdV has exact soliton solution \cite{Dey96,Bri02}, so we use this solution as the initial condition in the example presented in Fig.~\ref{Fi4}. The explicit form of this solution is
\begin{equation} \label{5OrI}
\eta(x,t) = A\, \text{Sech}^4[B(x-vt)],
\end{equation}
where $A,B,v$ are functions of the coefficients of the 5th-order KdV equation (\ref{eWS2}).
For illustration we chose $\tau=0.35>\frac{1}{3}$ which assures $B\in \mathbb{R}$. 
The numerical evolution of the soliton (\ref{5OrI}) according to the equation (\ref{5kdvQ}) for $\delta=0$ (flat bottom) confirms that it moves with the constant shape and constant velocity.

Snapshots of numerical evolution of the fifth order KdV soliton (\ref{5OrI}) according to the equation (\ref{5kdvQ}) for $\delta=0.15$, $\alpha=0.2424, ~\beta=0.3$ are displayed in Fig.~\ref{Fi4}. In dimensionless coordinates this setup is the same as in previously discussed examples in subsections \ref{cas1n} and \ref{cas2n}. Coming back to dimension variables we realize that it is the system of completely different scale. First, taking $T=72\, mN/m$ as water surface tension one obtains water depth $
h=\sqrt{\frac{T}{\varrho g \tau}} \approx 0.0046\, m$. Next, with $\beta=0.3$ the dimensionless interval $x\in[0,15]$ corresponds to $[0, 0.123\,m]$. Indeed, the equation (\ref{5kdvQ}) describes soliton motion in capilary-gravity case with the uneven bottom. 
In general, the properties of this motion in dimensionless variables are similar to those observed in subsections \ref{cas1n} and \ref{cas2n} for the cases 1 and 2.

\subsection{Brief comparison} \label{comp}

In all three cases displayed in Figs.\ \ref{Fi2}, \ref{Fi3} and \ref{Fi4} the solitons move initially over the flat bottom with undisturbed shapes and constant initial velocities. Next, moving over the slope all solitons experience an amplitude increase and a corresponding decrease of velocity. 
The deformation of the profile, that is a lowering of the water level behind the soliton when it reaches the flat region is very small in Case 1 and 2. In Case 3 it is almost impercetible. 

\begin{figure}[hbt] 
\begin{center}
 \resizebox{0.8\columnwidth}{!}{\includegraphics[angle=270]{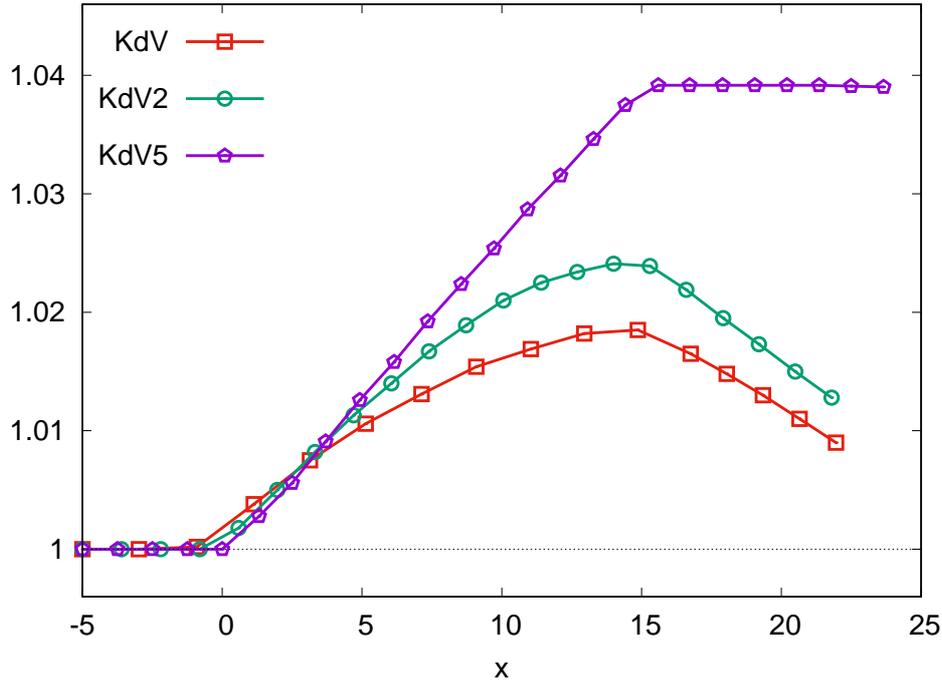}}
 \caption{
Solitons' maxima versus their positions from Fig.\ \ref{Fi2} - red symbols,  from Fig.\ \ref{Fi3} - green symbols and from Fig.\ \ref{Fi4} - blue symbols. In the case of KdV5 displayed are the rescaled absolute values.} \label{Fi5}
\end{center} 
 \end{figure}

Qualitatively, the properties of soliton motion calculated in nondimensional variables are very similar in all these three cases.  Particularly similar are cases 1 and 2. The differences are seen in detail in Fig.\ \ref{Fi5}
in which the ratios of solitons maxima to their initial values $\text{max}[\eta]/\eta_0$ versus their positions (read from the calculated data) are plotted. Since in KdV5 case the amplitudes are negative we plot their absolute values multiplied by $1/|A(t=0)|$ to compare the changes with respect to initial values.

More detail studies of numerical evolution according to the derived equations are planned in a near future.

\section{Conclusions} \label{concl}

In attempts to derive KdV-type equations for the case of the uneven bottom, 
we found two universal features.  
First, the boundary condition at the uneven bottom implies the limit on the order of theory in which the Boussinesq equations can be derived. It usually does not exceed the second order, whereas for the flat bottom one can proceed to arbitrary order.

Second, regardless of the different ordering of small parameters $\alpha,\beta,\delta$, the condition of compatibility of Boussinesq's equations containing the bottom terms is always the same in the lowest order and given by (\ref{warunek}).

This condition cannot be satisfied for an arbitrary bottom profile $h(x)$.
However, the condition (\ref{warunek}) is satisfied when the bottom relief is given by an arbitrary piecewise linear function. For such bottom profiles, we derived four KdV-type equations which generalize equations known for the flat bottom case. The equation (\ref{kdvDo}) generalizes the usual KdV equation. In (\ref{kdvDo}) terms originating from the bottom are of first order. The equation (\ref{2abQ}) generalizes the extended KdV (KdV2) equation.
The equation (\ref{5kdvQ}) generalizes the fifth-order KdV (KdV5) equation whereas (\ref{GardH}) generalizes the Gardner equation. In the last three cases, terms induced by the piecewise linear bottom are of second order.\\

{\bf Acknowledgement}

P.R. thanks for the financial support from the program of the Polish Minister of Science and Higher Education under the name ”Regional Initiative of Excellence” in 2019 - 2022, project no. 003/RID/2018/19, funding amount 11 936 596.10 PLN.



\begin{thebibliography}{99}

\bibitem{kdv} Korteweg, D.J.\ and  de Vries, G.: On the change of form of the long waves advancing in a rectangular canal, and on a new type of stationary waves. 
Phil.\ Mag.\ (5), {\bf 39}, 422 (1895) 

\bibitem{Whit} Whitham, G.B.: {\em Linear and nonlinear waves},  John Wiley \& Sons, New York, (1974) 

\bibitem{DrJ} Drazin P.G.\ and Johnson,  R.S.:  {\em Solitons: An Introduction},  Cambridge University Press, Cambridge, (1989) 
 
\bibitem{Ablc} Ablowitz, M.J.\ and Clarkson, P.A.:  {\em  Solitons, Nonlinear Evolution Equations and Inverse Scattering}, Cambridge University Press, Cambridge, (1991)

\bibitem{Hir} Hirota, R.: {\em  The Direct Method in Soliton Theory}, Cambridge University Press, Cambridge, (2004), first published in Japanese (1992)

\bibitem{Rem} Remoissenet, M.: {\em  Waves Called Solitons: Concepts and Experiments}, Springer, Berlin, (1999)

\bibitem{InR} Infeld, E.\ and Rowlands, G.: {\em  Nonlinear Waves, Solitons and Chaos}, Cambridge University Press, Cambridge, (2000)

\bibitem{Osb} Osborne, A.R.: {\em Nonlinear ocean waves and the inverse scattering transform}, Academic Press, Amsterdam, (2010)

\bibitem{Mei} Mei, C.C.\ and Le M\'ehaut\'e, B.: Note on the equations of long waves over an uneven bottom. J.\ Geophys. Research, {\bf 71}, 393-400 (1966)

\bibitem{Grim70} Grimshaw, R.:  The solitary wave in water of variable depth.
J.\ Fluid Mech.  {\bf 42}, 639-656 (1970)

\bibitem{Djord} Djordjevi\'c V.D.\ and Redekopp, L.G.: On the development of packets of surface gravity waves moving over an uneven bottom. J.\ Appl.\ Math. and Phys.\ (ZAMP), {\bf 29}, 950-962 (1978) 

\bibitem{BH} Benilov, E.S.\ and Howlin,  C.P.: Evolution of Packets of Surface Gravity Waves over Strong Smooth Topography. Studies in Appl.\ Math., {\bf 116}, 289-301 (2006)

\bibitem{Pel} Nakoulima, O., Zahibo, N., Pelinovsky, E., Talipova, T.\ and  Kurkin, A.: Solitary wave dynamics in shallow water over periodic topography. Chaos, {\bf 15}, 037107 (2005)

\bibitem{Peli} Grimshaw, R., Pelinovsky E.\ and Talipova, T.\: Fission of a weakly nonlinear interfacial solitary wave at a step.  Geophys. Astrophys. Fluid Dynamics,  {\bf 102}, 179-194 (2008)\

\bibitem{Peli1}
E.\ Pelinovsky, B.H.\ Choi, T.\ Talipova, S.B.\ Woo and D.C.\ Kim,  \emph{Solitary wave transformation on the underwater step: theory and numerical experiments}, 
 Appl. Math.\ Comput.\,  {\bf 217}, 1704-1718 (2010).

\bibitem{Grim} Grimshaw, R.H.J.\ and  Smyth, N.F.: Resonant flow of a stratified fluid over topography. J.\ Fluid.\ Mech. {\bf 169} 429-464 (1986)

\bibitem{Smy} Smyth, N.F.: Modulation Theory Solution for Resonant Flow Over Topography. Proc.\ R.\ Soc.\ Lond.\ A,  {\bf 409}, 79-97 (1987)

\bibitem{PS98} Pelinovskii, E.N, and Slyunayev, A.V.: Generation and interaction of large-amplitude solitons. JETP Lett. \textbf{67}, 655-661 (1998)

\bibitem{Kam}
Kamchatnov, A.M., Kuo, Y.-H., Lin, T.-C., Horng, T.-L., Gou, S.-C., Clift, R., El, G.A.\ and Grimshaw, R.H.J.: Undular bore theory for the Gardner equation. Phys.\ Rev.\ E {\bf 86},  036605 (2012)
 

\bibitem{G&P1} van Greoesen, E.\ and Pudjaprasetya, S.R.: Uni-directional waves over slowly varying bottom. Part I: Derivation of a KdV-type of equation. 
Wave Motion, {\bf 18}, 345-370 (1993)
 
\bibitem{G&P2}  Pudjaprasetya, S.R.\ and van Greoesen, E.: Uni-directional waves over slowly varying bottom. Part II: Quasi-homogeneous approximation of distorting waves. Wave Motion, {\bf 23}, 23-38 (1996)

\bibitem{GN} Green A.E.\ and Naghdi, P.M.: A derivation of equations for wave propagation in water of variable depth. J.\ Fluid Mech., {\bf 78}, 237-246 (1976)

\bibitem{Kim} J.W.\ Kim,  J.W., Bai, K.J., Ertekin R.C.\ and Webster, W.C.: A derivation of the Green-Naghdi equations for irrotational flows.  
J.\ Eng.\ Math., {\bf 40}, 17-42 (2001)

\bibitem{Nad}
 Nadiga, B.T., Margolin, L.G.\ and Smolarkiewicz, P.K.: Different approximations of shallow fluid flow over an obstacle. Phys. Fluids,   {\bf 8}, 2066-2077 (1996)
 
\bibitem{Sel2000} Selezov, I.T.: Propagation of unsteady nonlinear surface gravity waves above an irregular bottom. International Journal of Fluid Mechanics.
 \textbf{27} (1), 146-157 (2000)

\bibitem{NiuYu} Niu, X.\ and Yu X.: Analytic solution  of long wave propagation over a submerged hump. Coastal Engineering, {\bf 58}, 143-150 (2011); -  
Liu, H-W.\ and Xie, J-J.: Discussion of "Analytic solution  of long wave propagation over a submerged hump" by Niu and Yu (2011). Coastal Engineering, {\bf 58},) 948-952 (2011)


\bibitem{YGJC18} Yuan, C., Grimshaw, R., Johnson, E.\ and Chen, X.: 
The Propagation of Internal Solitary Waves over Variable Topography in a Horizontally Two-Dimensional Framework. Journal of Physical Oceanography 
\textbf{48}, 283-300 (2018)
 https://doi.org/10.1175/JPO-D-17-0154.1

\bibitem{FY2018} Fan, L., Yan, W.: On the weak solutions and persistence properties for the variable depth KDV general equations. Nonlinear Analysis: Real World Applications \textbf{44} 223–245, (2018)
https://doi.org/10.1016/j.nonrwa.2018.05.002

\bibitem{YS2018} Stepanyants, Y.: The effects of interplay between the rotation and shoaling for a solitary wave on variable topography. Studies in Applied Mathematics. 2019; 1-22. https://doi.org/10.1111/sapm.12255

\bibitem{KRR} Karczewska, A., Rozmej, P. and Rutkowski, \L{}.:
A new nonlinear equation in the shallow water wave problem. 
Physica Scripta, \textbf{89}, 054026, (2014).  
https://doi.org/10.1088/0031-8949/89/5/054026

\bibitem{KRI} Karczewska, A., Rozmej, P.\ and Infeld, E.:
Shallow water soliton dynamics beyond KdV.
Phys.\ Rev.\ E, \textbf{90}, 012907, (2014) 
https://doi.org/10.1103/PhysRevE.90.012907

\bibitem{BurSerg} Burde, G.I.\ and Sergyeyev, A.:
Ordering of two small parameters in the shallow water wave problem. 
J. Phys. A: Math. Theor. \textbf{46},  075501, (2013)

\bibitem{Burde} The review by the anonymous referee of the paper:
 P.~Rozmej, A.~Karczewska, \emph{Comment on the paper "The third-order perturbed Korteweg-de Vries equation for shallow water waves with a non-flat bottom" by M. Fokou, T.C. Kofané, A. Mohamadou and E. Yomba, Eur. Phys. J. Plus, 132, 410 (2017)}. arXiv:1804.01940.




\bibitem{MS90} Marchant, T.R.\ and Smyth, N.F.:
The extended Korteweg-de Vries equation and the resonant flow
of a fluid over topography. Journal of Fluid Mechanics, \textbf{221}, 263-288, (1990)

\bibitem{IKRR}  Infeld, E., Karczewska, A., Rowlands, G.\ and Rozmej, P.:  Exact cnoidal solutions of the extended KdV equation. Acta Phys. Pol. A, \textbf{133}, 1191-1199, (2018) 
https://doi.org/10.12693/APhysPolA.133.1191

\bibitem{RKI} Rozmej, P., Karczewska, A.\ and Infeld, E.: Superposition solutions to the extended KdV equation for water surface waves. Nonlinear Dynamics  \textbf{91}, 1085-1093, (2018) 
https://doi.org/10.1007/s11071-017-3931-1

\bibitem{RK} Rozmej, P.\ and Karczewska, A.:
New Exact Superposition Solutions to KdV2 Equation.  Advances in Mathematical Physics. \textbf{2018}, Article ID 5095482, 1-9, (2018)  
https://doi.org/10.1155/2018/5095482


\bibitem{HS88} Hunter, J.K.\ and Scheurle, J.: Existence of perturbed solitary wave solutions to a model equation for water waves. Physica D. \textbf{32}, 253-268, (1988).

\bibitem{GMB94} Grimshaw, R., Malomed, B.\ and Benilov, E.: Solitary waves with damped oscillatory tails -- an analysis of the 5th-order Korteweg-de Vries equation. Physica D. \textbf{77}, 473-485, (1994)

\bibitem{Dey96} Dey, B., Khare, A. and Kumar, C.N.: Stationary solitons of the fifth order KdV-type. Equations and their stabilization. Phys. Lett. A.
{\bf 223}, 449-452 (1996)

\bibitem{Bri02} Bridges, T.J., Derks, G. and Gottwald, G.: Stability and instability of solitary waves of the fifth order KdV equation: a numerical framework. Physica D. {\bf 172}, 190-216 (2002)

\bibitem{RH93} Rosenau, P.\ and Hyman, J.M.: Compactons: solitons with finite wavelength. Phys.\ Rev.\ Lett. \textbf{70}, 564-567 (1993)

\bibitem{KY78} Kakutani, T.\ and Yamasaki, T.: Solitary waves on a two-layer fluid. J. Proc. Soc. Japan, \textbf{45}, 674-679 (1978)


\end{thebibliography}
\end{document}